# Giant enhancement of critical current density at high field in superconducting (Li,Fe)OHFeSe films by Mn doping

Dong Li[1,2], Jie Yuan[1,2], Peipei Shen[1,2], Chuanying Xi[4], Jinpeng Tian[1,2], Shunli Ni[1,2], Jingsong Zhang[1,2], Zhongxu Wei[1,2], Wei Hu[1,2], Zian Li[1,2], Li Yu[1,3], Jun Miao[5], Fang Zhou[1,2], Li Pi[4], Kui jin[1,2,3,], Xiaoli Dong[1,2,3, a)], and Zhongxian Zhao[1,2,3]

[1]Beijing National Laboratory for Condensed Matter Physics, Institute of Physics, Chinese Academy of Sciences, Beijing 100190, China.

[2]School of Physical Sciences, University of Chinese Academy of Sciences, Beijing 100049, China.

[3]Songshan Lake Materials Laboratory, Dongguan, Guangdong 523808, China.

[4]Anhui Province Key Laboratory of Condensed Matter Physics at Extreme Conditions, High Magnetic Field Laboratory of the Chinese Academy of Sciences, Hefei 230031, Anhui, China.

[5]Beijing Advanced Innovation Center for Materials Genome Engineering, School of Materials Science and Engineering, University of Science and Technology Beijing, Beijing 100083, China.

[a)] dong@iphy.ac.cn

**Abstract:**

**Critical current density ($J_c$) is one of the major limiting factors for high field applications of iron-based superconductors. Here, we report that Mn-ions are successfully incorporated into nontoxic superconducting (Li,Fe)OHFeSe films. Remarkably, the $J_c$ is significantly enhanced from 0.03 to 0.32 MA/cm² under 33 T, and the vortex pinning force density monotonically increases up to 106 GN/m³, which is the highest record so far among all iron-based superconductors. Our results demonstrate that Mn incorporation is an effective method to optimize the performance of (Li,Fe)OHFeSe films, offering a promising candidate for high-field applications.**

## Introduction

Critical current density, the maximal ability of superconductors to carry current without dissipation, is a crucial factor for high-field applications[1,2]. Among all superconductors, the record $J_c$ value is held so far by the copper oxide superconductors, but their practical application is hampered by a few obstacles[2], such as high anisotropy, a small critical grain boundary angle ($\theta_c$), and a high production cost. On the other hand, iron-based superconductors have moderate anisotropy, a high irreversibility field ($H_{irr}$), and $\theta_c$, making them more promising for high field application[2-10]. There are still drawbacks for the high-field application of iron-based superconductor. For example, the toxic element arsenic in iron pnictides limits their application although their transition temperature ($T_c$) and $J_c$ are relatively high[2-5,9]. Therefore, it further research should aim to find nontoxic iron-based superconductors with comparable or even higher $J_c$.

The nontoxic newly discovered superconductor (Li,Fe)OHFeSe (FeSe-11111)[11], with an optimal $T_c$ of 42 K and a self-field $J_c$ of 0.5 MA/cm² at 20 K under ambient pressure[12], turns out to be a good candidate. However, the vortex pinning potential of FeSe-11111 is relatively low due to the large layer distance[13], leading to the broadening of resistive transition under a magnetic field[14,15]. This calls for further efforts at improving the vortex pinning ability of FeSe-11111, for instance by embedding extra vortex pinning defects[16-19]. Recently, elemental Mn has been incorporated into FeSe-11111 single crystals without obvious detriment to its $T_c$[20], which may provide an effective candidate. Moreover, iron-based superconductors in the form of films usually present a higher $J_c$ than that of bulk samples[6,8]. Therefore, it is worthy introducing transition metal ions into FeSe-11111 crystalline superconducting film for further optimization of their high-field performance.

In this letter, we successfully introduced Mn-ions into a superconducting FeSe-11111 film synthesized through the so-called matrix-assisted hydrothermal epitaxy (MHE) method[12]. A significant enhancement of $J_c$ was observed in FeSe-11111 films by Mn-doping, increasing it tenfold from 0.03 to 0.32 MA/cm² under 33 T at 5 K. Remarkably, the vortex pinning force density ($F_p$) of Mn-doped films monotonically increases to 106 GN/m³. To the best of our knowledge, this is the highest record so far among all iron-based superconducting systems. By analyzing $F_p$ versus magnetic fields, we find the apparent enhancement of $J_c$ in the Mn-doped FeSe-11111 film stems from the extra pinning centers induced by Mn doping.

## Experiments

The pure and Mn-doped FeSe-11111 films were synthesized via the MHE method that we developed[12]. The x-ray diffraction (XRD) experiments were carried out on a 9 kW Rigaku SmartLab X-ray diffractometer. The scanning electron microscope (SEM) and energy dispersive X-ray (EDX) spectroscopy measurements were performed on a Hitachi SU5000. The electron energy loss spectroscopy (EELS) data were acquired using a transmission electron microscope (ARM200F, JEOL Inc.) equipped with a Gatan Quantum ER 965 Imaging Filter. Electrical transport measurements within 9T were collected with the standard four-probe method on a Quantum Design PPMS-9 system. The values of $J_c$ were obtained using the criteria of 1 μV on I-V curves and the parameters of bridge were characterized by SEM. The high-field experiments up to 33 T were performed on the Steady High Magnetic Field Facilities, High Magnetic Field Laboratory, CAS.

## Results and discussion

### Characterization of crystal structure and elemental Mn

Figure 1(a) shows the XRD patterns of Mn-doped (top) and Mn-free (bottom) FeSe-11111 films, respectively. There are no detectable impurity phases in the Mn-doped system. Moreover, both of them exhibit a single preferred orientation of (00l) and the peaks of LaAlO₃ substrates are marked with LAO. The calculated c-axis length decreases from 9.33 Å of the Mn-free film to 9.30 Å of the Mn-doped film, consistent with the result from single crystals[21]. Albeit the full width at half maximum (FWHM) of the rocking curve for the (006) reflection expands from 0.15° to 0.38° due to the presence of Mn-ions, it is still smaller than most of other iron-based superconductor films[22-24], suggesting high crystalline quality. To verify the incorporation of elemental Mn, all the Mn-doped films were checked by EDX spectroscopy or EELS. One of the in-plane SEM images and corresponding EDX mapping of Mn-doped films are shown in Figure 1(b), which display smooth morphology and the uniform distribution of Mn, Fe, and Se elements. The atomic ratio of Mn:Se was determined to be 0.12 by EDX spectroscopy. Moreover, one typical EELS pattern of Mn doped films is shown in Figure 1(c). The presence of $L_{3,2}$ edges for Mn and Fe indicates that Mn ions are incorporated into the lattice of FeSe-11111 films, not left as impurities, which is consistent with the case of Mn-doped FeSe-11111 single crystals[21]. Because of the small difference in ionic sizes for Mn and Fe, it is expected that the incorporated Mn ions substitute for Fe ions at the crystallographic tetrahedral sites. Now that the homogeneous incorporation of Mn-ions has been confirmed in the Mn-doped films, we can make a comparative study on the electrical transport properties between Mn-doped and Mn-free FeSe-11111 films.

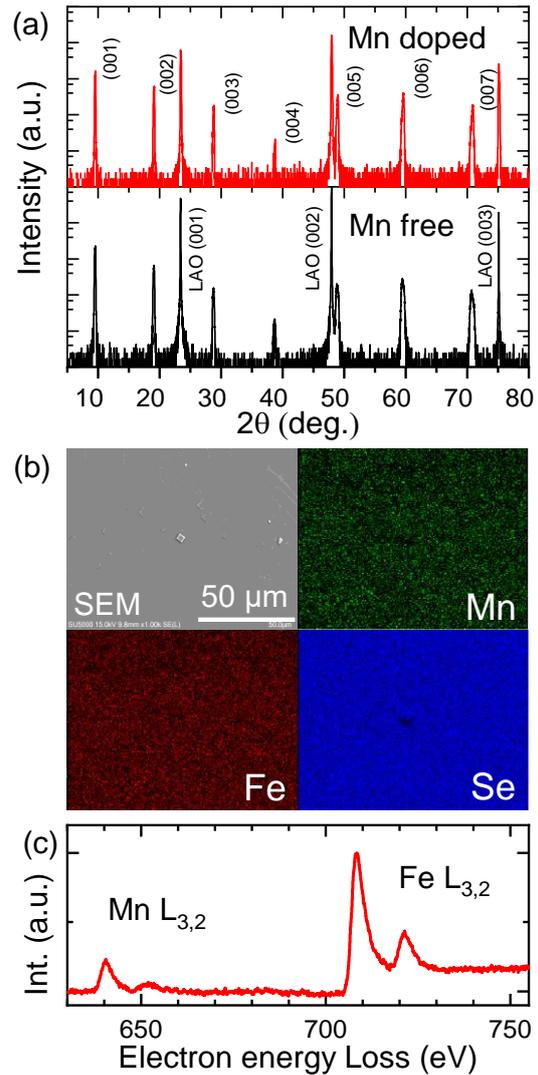

**Figure 1.** (a) XRD patterns of Mn-doped and Mn-free superconducting FeSe-11111 films, respectively. (b) SEM image and corresponding EDX element mapping of Mn, Fe and Se on the Mn-doped film. (c) The EELS data of Mn-doped FeSe-11111 films.

*Electrical transport properties*

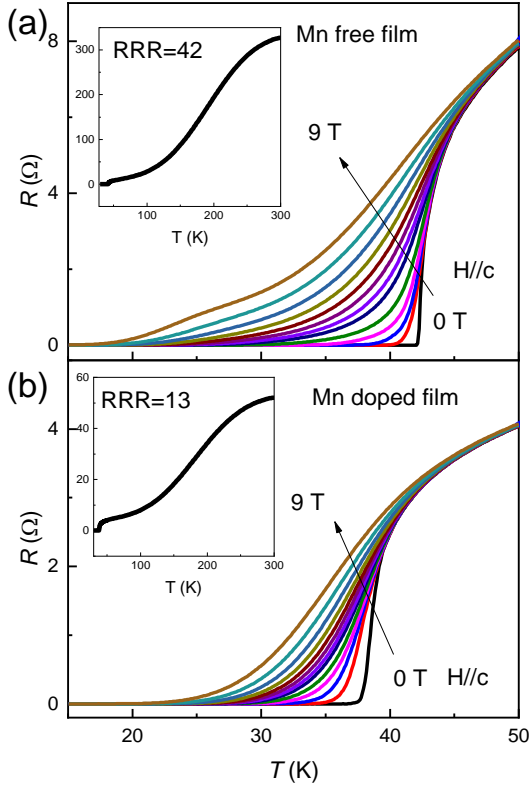

**Figure 2.** (a) and (b) Temperature dependent resistance of Mn-free and Mn-doped FeSe-11111 films under various c-axis magnetic fields up to 9T, respectively. The insets show the RRR = R$_{300\,K}$/R$_{50\,K}$ for corresponding films.

The temperature-dependent resistance (R-T) of pure and Mn-doped FeSe-11111 films under *c*-axis magnetic fields are shown in Figures 2(a) and 2(b), respectively. As elemental Mn is incorporated, the $T_c$, the onset temperature of zero-resistance, declines from 42.0 K to 36.6 K and the residual resistance ratio (RRR = R$_{300\,K}$/R$_{50\,K}$, insets of Figures 2(a) and 2(b)) decreases from 42 to 13, which indicates that the Mn ions weakly reduce the superconductivity and serve as extra pairing breaking centers. With increasing fields, the resistive transition becomes broadened and tailed, as widely observed in high-$T_c$ cuprates and iron-based superconductors[25-27], and it is more evident in Mn-free films. This feature is caused by the thermally assisted flux flow and reflects the strength of vortex pinning force[27]. Therefore, the ability of vortex pinning in FeSe-11111 films is indeed improved by Mn doping.

Figure 3(a) displays the representative I-V curves of Mn-doped films, from which the $J_c$ is extracted using the criterion of 1 μV. Figure 3(b) shows the temperature dependence of $J_c$ for Mn-doped (red) and Mn-free (black) films under magnetic fields along (open symbols) or perpendicular (closed symbols) to c-axis. The measurable ranges of $J_c$ were limited by the size of bridges and the upper limit of the applied current, leading to the absence of high $J_c$ data hereafter. Due to the intrinsic electronic two-dimensional (2D) property of FeSe-11111[14], the out-of-plane magnetic field suppresses the superconductivity more prominently that in-plane one, accounting for the significant $J_c$ anisotropy. It is noted that the $J_c$ values of both Mn-free and Mn-doped films maintain the order of $1\times10^5$ A/cm$^2$ under 9 T at 10 K, which is higher than the value of $1\times10^2$ A/cm$^2$ in FeSe-11111 single crystals[28]. The depairing current density of (Li,Fe)OHFeSe can be estimated by $J_0 = \frac{\varphi_0}{3\sqrt{3}\mu_0\lambda_c^2\xi_0^{ab}}$, where $\varphi_0$ is the flux quantum, and using values of the penetration depth $\lambda \sim 281$ nm[29] and the coherence length $\xi^{ab} \sim 2.21$ nm[15]. The calculated value of $J_0 \sim 57.7$ MA/cm$^2$ indicates that the FeSe-11111 material has great potential of $J_c$. Thus, such a dramatic enhancement of $J_c$ is an intrinsic property of the high-quality films [12,30]. It demonstrates that superconducting FeSe-11111 crystalline films are superior to single crystals to achieve high $J_c$, as commonly seen in other iron-based superconductors[6,8]. Moreover, it is apparent that the enhancement of $J_c$ in FeSe-11111 films by Mn doping under both out-of-plane and in-plane magnetic fields. For instance, the resultant value of $J_c$ increases from 0.9 to $5.1\times10^5$ A/cm$^2$ under 9 T (*H // c*) at 10 K. Even for the self- field $J_c$, the performance of Mn-doped films is superior to Mn-free films below 30 K. In addition, the enhancement of $J_c$ via Mn doping keeps increasing with a decreasing temperature, making the Mn-doped films more interesting in the low-temperature, high-field magnet application. This result confirms that Mn doping is beneficial to the enhancement of $J_c$ for FeSe-11111 films.

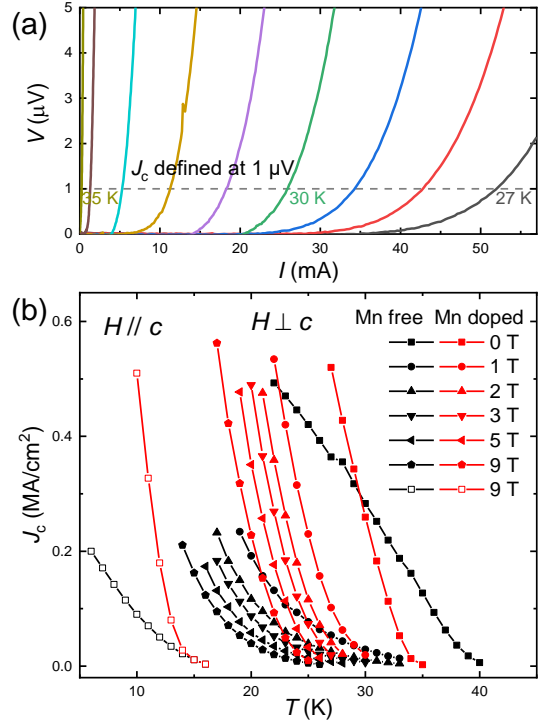

**Figure 3.** (a) The representative *I-V* curves of Mn-doped films measured from 35 K to 27 K under 0 T. (b) Temperature dependence of $J_c$ for the Mn-doped (red) and Mn-free (black) films under various applied fields along in-plane (H⊥c, closed symbol) and out-of-plane (H // c, open symbol), respectively.

*Pinning mechanism analysis*

To elucidate the origin of $J_c$ enhancement in Mn-doped FeSe-11111 films, we have investigated the magnetic field dependence of $F_P$ (=$J_c$ × $H$). Due to the strong tail effects of R-T curves, it is difficult to determine the $H_{irr}$ accurately, which is commonly used to normalize the magnetic field[31,32]. In order to improve the accuracy, the normalized pinning force density f=$F_P$/$F_{p,max}$ is scaled with field h = $H$/$H_{max}$, where the $F_{p,max}$ is the maximal $F_p$ and $H_{max}$ is the corresponding magnetic field [33,34]. Figure 4 shows plots of scaled data for Mn-doped (red) and Mn-free (black) films. They both follow a temperature independent scaling, indicating there is single dominant pinning mechanism within measured temperature range. Thus, we can analyze the scaling of $f$ (h) by the following equations[18]:

$$f = \frac{9}{4} h \left(1 - \frac{h}{3}\right)^2 \qquad \text{for normal point pinning,} \quad (1)$$

$$f = \frac{25}{16} h^{0.5} \left(1 - \frac{h}{5}\right)^2 \qquad \text{for normal surface pinning.} \quad (2)$$

It turns out that both cases can be well fitted by equation (2) below $H_{max}$, demonstrating the dominant pinning mechanism owing to normal surface pinning. These equations actually originate from the Dew-Huges model[32], where the normal surface pinning refers the pinning centers arising from 2D non-superconducting area such as grain boundaries, plate-like precipitates, and surface of superconductors. This outcome is reasonable given that the insulating (Li,Fe)OH spacer layers may serve as the surface pinning centers in FeSe-11111 superconductors. Meanwhile, above $H_{max}$, both scaled plots deviate from the theoretical fitting curve, and this deviation is more pronounced in the case of Mn-doped films. The similar feature was also observed in YBa₂Cu₃O₇₋δ [33,34], Bi₂Sr₂CaCu₂O₈₊ₓ[35], Nd₂₋ₓCeₓCuO₄₊δ[36], and Mn doped KₓFe₂₋ᵧSe₂ single crystals[18], which were attributed to the effect of flux creep. In the normal surface pinning, the $F_p$ is proportional to the pinning center density, implying that the improvement of $J_c$ stems from the extra pinning centers induced by Mn doping. This is consistent with the prior evidence of the expanded FWHM of the rocking curve and the decreased RRR value in Mn-doped films, due to the increase of crystal defects and scattering centers. Further investigation is needed into how the Mn ions affect the vortex behavior of FeSe-11111 films. One possibility we can speculate on is that the Mn ions are doped partly into the (Li,Fe)OH- interlayers, with the point defect concentration in the FeSe-layers not sufficient to change the surface pinning behavior.

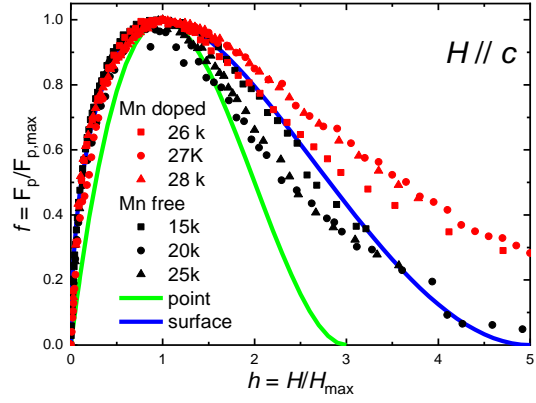

**Figure 4.** Normalized vortex pinning force density f = $F_p$/$F_{p,max}$ versus field h = $H$/$H_{max}$ at various temperature for Mn-doped and Mn-free films. Green line is the fit curve of normal point pinning and the blue line is related to normal surface pinning.

*High-field performance*

Figures 5(a) and 5(b) present the data of high-field (*H // c*) $J_c$ and $F_p$ for Mn-doped and pure FeSe-11111 films at 5 K. Data of SmFeAsO(O,F) films[37], FeSe₀.₅Te₀.₅ films[7], P-doped BaFe₂As₂ films[38] and YBa₂Cu₃O₇₋δ wires[39] at 4.2 K were also included for comparison. The enhancement of FeSe-11111 films, particularly for the high-field performance, is significant due to Mn doping. Notably, the high field tolerance of $J_c$ for the Mn-doped film overwhelms other iron-based superconductors, which is essential for high-field magnet applications. Numerically, the corresponding $J_c$ increases from 0.03 to 0.32 MA/cm² under 33 T, three times as large as 0.1 MA/cm², the widely accepted value for practical application[2,9]. It is worthy to note that the $F_p$ of the Mn-doped film monotonically increases up to 106 GN/m³, as shown in Figure 5(b), and it is the only iron-based superconductor that resembles the high-field performance of YBa₂Cu₃O₇₋δ to date. This feature indicates that the Mn-doped FeSe-11111 films provide a potential extra material for extremely high magnetic field applications, besides cuprates[40]. To our knowledge, the high-field $J_c$ and $F_p$ of Mn-doped FeSe-11111 films obtained in this work set new records so far among all iron-based superconductors. Moreover, the $J_c$ of Mn-doped films may increase further given that: 1) the $T_c$ of Mn-doped (Li,Fe)OHFeSe superconductors could be further tuned up to 41 K with proper Mn incorporation[21] and 2) introducing extra artificial pinning centers, similar to proton or ion irradiation, could enhance $J_c$ more significantly[17,19].

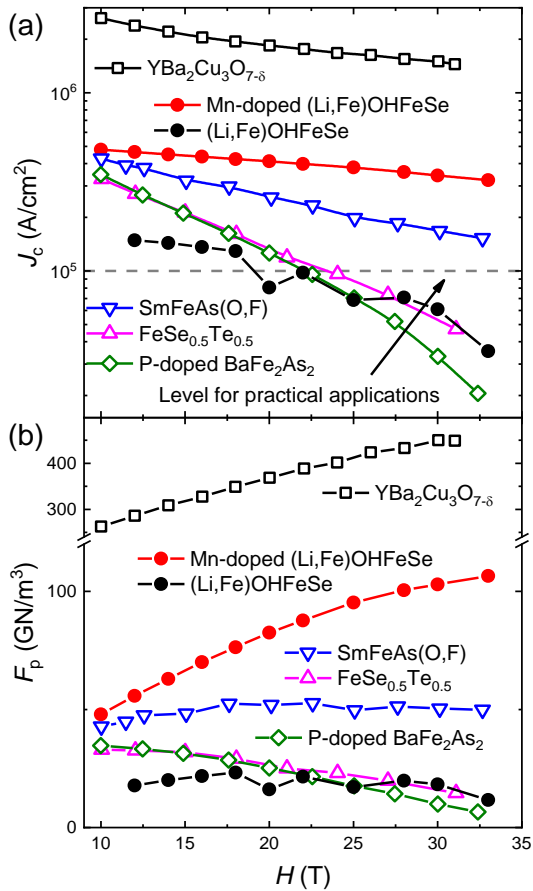

**Figure 5.** Magnetic field dependence of $J_c$ (a) and $F_p$ (b) of several superconductors, including Mn-doped and pure (Li,Fe)OHFeSe films at 5 K, SmFeAs(O,F) films [37], FeSe$_{0.5}$Te$_{0.5}$ films [7], P-doped BaFe$_2$As$_2$ films[38], and YBa$_2$Cu$_3$O$_{7-\delta}$ wires[39] at 4.2 K under c-axis fields.

## Conclusion

The Mn ions are uniformly incorporated into superconducting FeSe-11111 crystalline films and $J_c$ is remarkably enhanced to 0.32 MA/cm$^2$ under 33 T, which is, to our knowledge, the highest reported value so far among various iron-based superconductors. It is found that the normal surface pinning dominates both films and the giant enhancement of $J_c$ stems from extra pinning centers induced by Mn doping. Combined with the energy-saving hydrothermal synthesis, Mn doping method utilized in this work can serve as an easy, fast, and low-cost way to effectively enhance $J_c$ for intercalated iron selenide superconductors. These results demonstrate that the nontoxic Mn-doped superconducting (Li,Fe)OHFeSe crystalline films have very promising prospects of high-field applications.

## Acknowledgments

We thank Z. F. Lin and S. S. Yue for valuable discussions. This work was supported by the National Key Research and Development Program of China (Grant Nos. 2017YFA0303003, 2016YFA0300301, 2017YFA0302902, 2018YFB0704102), the National Natural Science Foundation of China (Nos.11888101, 11834016, 11674374, 11574027), the Strategic Priority Research Program of Chinese Academy of Sciences (XDB25000000), the Strategic Priority Research Program and Key Research Program of Frontier Sciences of the Chinese Academy of Sciences (Grant Nos. QYZDY-SSW-SLH001, QYZDY-SSW-SLH008)